\documentclass[preprint2]{aastex}
\usepackage{graphicx, rotating}
\DeclareGraphicsExtensions{.eps}

\shorttitle{W40}
\shortauthors{Rodney}

\begin{document}

\title{A Cluster of Compact Radio Sources in W40}

\author{Luis F. Rodr{\'{\i}}guez}
\affil{Centro de Radioastronom\'ia y Astrof\'isica, UNAM\\
 Apdo. Postal 3-72 (Xangari),58089 Morelia, Michoac\'an, M\'exico}
\email{l.rodriguez@crya.unam.mx}

\author{Steven A. Rodney}
\affil{The Johns Hopkins University\\
        3400 N. Charles St., Baltimore, MD 21210 USA}
\email{rodney@jhu.edu}

\author{Bo Reipurth}
\affil{Institute for Astronomy, University of Hawaii at Manoa \\
        640 N. Aohoku Place, Hilo, HI 96720, USA}
\email{reipurth@ifa.hawaii.edu}

\begin{abstract}
 We present deep 3.6 cm radio continuum observations of the HII region W40
 obtained using the Very Large Array in its A and B configurations.  We
 detect a total of 20 compact radio sources in a region of 4$\arcmin \times $
 4$\arcmin$, with 11 of them concentrated in a band with 30$\arcsec$ of
 extent.  We also present $JHK$ photometry of the W40 cluster taken with the
 QUIRC instrument on the University of Hawaii 2.2 meter telescope. 
 These data reveal that 15 of the 20 VLA sources have infrared counterparts,
 and 10 show radio variability with periods less than 20 days.
 Based on these combined radio and IR data, we propose that 8 of the
 radio sources are candidate ultracompact HII regions, 7 are likely to be
 young stellar objects, and 2 may be shocked interstellar gas. 
 
\vspace{2mm}
\vspace{6mm}
\end{abstract}

\section{Introduction} The W40 HII region (denoted as S64 in the Sharpless
catalog) is the central object in a rich
cloud complex that lies some $3.5\degr$ above the Galactic plane at J2000
coordinates $18^h 31^m 29^s$, $-02\degr 05\arcmin 4\arcsec$
\citep{wes1958,sha1959}. W40 is a blister HII region with a diameter of
$\sim 6\arcmin$. It is bordered by a molecular cloud that subtends an angle
of $\sim 1\degr$.  The dense core of this molecular cloud is 
located at galactic coordinates 28.77 +3.70, and is labeled TGU~279-P7 in the
all-sky cloud atlas of \citet{dob2005}. 

W40 has been well-studied at radio and millimeter
wavelengths, leading to a firm understanding of the region's gaseous
structures.
\citet{cru1982} compiled a comprehensive review of the available molecular line
and radio continuum data to develop a detailed kinematic picture of the HII
region and its interaction with the molecular cloud.
The warm CII interface between these two principal components of the cloud
complex has been extensively studied as well \citep{val1987,val1991,val1992}.
Additionally, Zeeman
measurements of OH absorption lines have revealed a magnetic field of
B=$-14.0\pm 2.6 \mu G$ \citep{cru1987}.

A dense stellar cluster within W40 is apparent in the IR maps of the 2MASS
and DENIS surveys \citep{rey2002}.  This cluster is dominated by
three central OB stars that constitute the primary excitation sources of the
HII region \citep{zei1978,smi1985}. The stars in this cluster appear to be
heavily obscured along our line of sight, with $A_{V} \sim 10$ magnitudes of
visual extinction \citep{shu1999}. W40 IRS2a has been suggested as the dominant
source of ionizing radiation in this region \citep{smi1985}, and millimeter
observations have shown evidence for  circumstellar dust shells around IRS1a,
2a, and 3a \citep{smi1985,val1994}. 

From the collective results of these studies we can see that the W40 HII region
has broken through the surrounding molecular cloud in a direction that lies
at an angle 
to our line of sight.  Thus, the stellar cluster which powers the HII region
is mostly hidden in the optical and near-IR, but scattered H$\alpha$ emission
marks the blistering edge of the HII region. 
The dense core of the molecular
cloud is offset from the HII region and its embedded cluster by $\sim
2\arcmin$, and a thin CII region delineates the boundary between the cool
molecular gas and the hot ionized region. 

The distance to W40 is not yet well constrained.  Several measurements of
atomic and molecular lines along the line of sight towards W40
have been used to estimate the distance to W40 $-$ assuming that the cloud
complex is in a circular orbit around the galactic center \citep{rei1970,dow1970,rad1972}.  
Taken together, these measurements produce only a weak constraint of 300-900
pc. Using assumptions about the spectral types of several of the stars in the
W40 cluster, distance modulus estimates can only limit the distance to a
wide range from 400 to 700 pc \citep{cru1982,smi1985}.  \citet{kol1983}
calculated a distance of 600 pc using a novel distance determination
technique involving OH line measurements. Hereinafter we will adopt a
distance of 600 pc, which puts the W40 complex at a height of 37 pc above the
galactic plane. 

More recently, the properties of the W40 region have been summarized
in a review by \citet{rod2008}. Also, \citet{kuh2010}
have performed a deep X-ray study of W40, and detected about 200 young
stellar members of the cluster.

\section{Observations}

We have used the Very Large Array (VLA) of the NRAO in its B and A
array configurations to observe W40 at 3.6 cm, under the NRAO program
code AR551.  The B configuration observations were made on 2003
November 3 while the A configuration observations were made on 2004
September 18.  Our phase center was at $\alpha (J2000) = 18^h 31^m
23\fs7$; $\delta (J2000) = -02\degr 05\arcmin 29\arcsec$.  The
amplitude calibrator was $1331+305$, with an adopted flux density of
5.18 Jy, and the phase calibrator was $1804+010$, with flux densities
of $0.76\pm 0.01$ and $0.82\pm 0.01$ Jy for the first and second
epochs of observation, respectively.  The half power contour of the
synthesized beams were $0\rlap.{''}88 \times 0\rlap.{''}76; +17^\circ$
(for configuration B data) and $0\rlap.{''}26 \times 0\rlap.{''}22;
+16^\circ$ (for configuration A data), for images made with the ROBUST
parameter of the task IMAGR set to 0 (Briggs 1995).  The data were
analyzed in the standard manner using the NRAO AIPS package. The data
for the two epochs were self-calibrated in phase separately and then
concatenated to obtain a final image with the highest sensitivity
possible.  The half power contour of the synthesized beam for this
concatenated data was $0\rlap.{''}36 \times 0\rlap.{''}30; +28^\circ$
for images made with the ROBUST parameter of the task IMAGR set to 0
\citep{bri1995}.

A total of 20 sources were detected in a field of $4\arcmin \times
4\arcmin$ in this image.  Following \citet{fom2002}, we estimate that
in a field of $4\arcmin \times 4\arcmin$ the {\em a priori} number of
expected 3.6 cm sources above $\sim$0.2 mJy is only $\sim$0.3. We then
conclude that perhaps one out of the 20 sources could be a background
object, but that we are justified in assuming that all the detected
sources are associated with W40.

Infrared photometry was carried out with the QUick Infrared Camera
(QUIRC, \citealt{hod1996}) on the University of Hawaii 2.2 meter
telescope from 2003 June 17 to 19. Data were obtained in the $J$, $H$,
and $K^\prime$ filters of the Mauna Kea Observatories (MKO) filter set
\citep{tok2005}.  These IR images were reduced and combined using the
Gemini Observatory's QUIRC package in IRAF. Our data were collected
under non-photometric conditions, and errors in the telescope guiding
combined with persistent focus problems made it impossible to pursue
photometry by fitting the point spread function, as is preferred for
crowded fields.  Thus, aperture photometry was performed using the
NOAO DIGIPHOT package. The four brightest IR stars were saturated even
in our shortest (1 sec) exposures, so without PSF fitting we were only
able to obtain upper limits on the magnitudes, which agree well with
previous work.

\begin{figure}[htbp]
\begin{center}
\includegraphics[width=\columnwidth,draft=false]{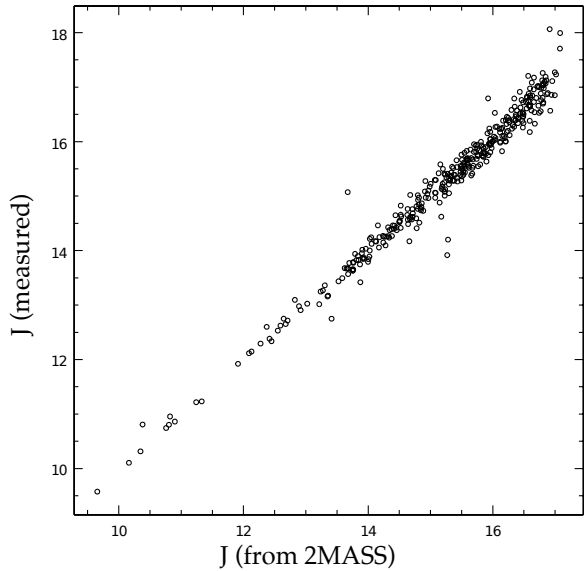}
\caption{Detected IR sources that appear in the 2MASS catalog, showing
  our measured $J$ magnitude plotted against the reported 2MASS $J$
  magnitude.  Both axes have been translated to the MKO filter
  system.}
\label{fig:errplots}
\end{center}
\end{figure}

We find over 1000 IR sources with detections in at least one of the
three bands, of which 271 have reliable photometric measurements in
all three bands.  A subset of 118 sources that have good $JHK^\prime$
photometry in our QUIRC data and good $JHK$ photometry in the 2MASS
catalog were used as calibration stars to translate the entire dataset
to the standard MKO system. This calibration step is the dominant
source of error in our IR photometry, because it relies on
transformation equations which convert our measured $K^\prime$ photometry into
the $K$ band and take the 2MASS $JHK$ magnitudes into the MKO $JHK$ system
(Connelley, private communication, \citealt{leg2006}). In Figure
\ref{fig:errplots} we demonstrate the scatter in our calibration step
by plotting the $J$ magnitudes we have measured and calibrated to the
MKO system against the $J$ magnitude reported by 2MASS, also translated
to the MKO system.  We estimate that this calibration procedure
introduces about 0.1 - 0.2 magnitudes of error into our final
photometry values.  A table of calibrated $JHK$ photometry for all 276
well-measured IR sources is available in the online edition of the
Astronomical Journal.

\begin{small}
\begin{deluxetable}{ccccc|ccc|c} 
\label{tab:vlasources}
\tablecolumns{9}
\tablecaption{Parameters of the 3.6 cm VLA Sources}
 
\tablehead{ \colhead{W40-VLA}
	  & \colhead{R.A.}
	  & \colhead{Dec}
	  & \colhead{Flux}
	  & \colhead{Time}
	  & \colhead{J}
	  & \colhead{H}
	  & \colhead{K}
	  & \colhead{W40-IRS} \\

 	    \colhead{Number}
	  & \colhead{$\alpha_{2000}$}
	  & \colhead{$\delta_{2000}$}
	  & \colhead{(mJy)}
	  & \colhead{Variable?}
	  & \multicolumn{3}{c}{(MKO)}
	  & \colhead{Designation\tablenotemark{*}} \\
	  }

\startdata

1  & 18 31 14.822 & -02 03 49.99 & 0.92 & N 
   &  8.22 & 7.507 & 6.98 \tablenotemark{\dag}
   & - \\  

2  & 18 31 15.279 & -02 04 15.19 & 0.90 & N 
   & - & - & -  
   & - \\  

3  & 18 31 22.328 & -02 06 19.60 & 0.47 & N 
   & 14.27 & 12.17 & 11.10 
   & - \\  

4  & 18 31 23.244 & -02 06 18.04 & 0.19 & Y($>$2.0) 
   & 14.27 & 12.72 & 11.72 
   & - \\  

5  & 18 31 23.622 & -02 05 35.80 & 3.97 & Y(5.3) 
   & 11.92 & 10.33 & 9.70 
   & - \\  

6  & 18 31 23.687 & -02 05 27.90 & 0.66 & Y($>$2.0) 
   & - & - & - 
   & \\  

7  & 18 31 23.985 & -02 05 29.53 & 2.40 & Y(3.2) 
   & 8.64 & 7.21 & 6.06 \tablenotemark{\dag}
   & IRS 2a \\  

8  & 18 31 26.021 & -02 05 17.02 & 3.27 & Y(8.3) 
   & 10.74 & 9.09 & 8.15
   & IRS 1c \\  

9  & 18 31 27.312 & -02 05 04.44 & 0.99 & Y(1.6) 
   & 13.51 & 12.03 & 11.17
   & - \\  

10 & 18 31 27.459 & -02 05 11.96 & 0.82 & Y($>$5) 
   & 11.41 & 9.68 & 8.79
   & - \\  

11 & 18 31 27.585 & -02 05 17.68 & 0.95 & Y(6.4) 
   & 13.65 & 12.18 & 11.20
   & - \\  

12 & 18 31 27.641 & -02 05 13.46 & 1.64 & N 
   & 13.50 & 11.68 & 10.47
   & - \\  

13 & 18 31 27.657 & -02 05 9.68 & 0.86 & Y($>$5) 
   & 10.57 & 9.36 & 8.59 \tablenotemark{\dag}
   & IRS 1d \\  

14 & 18 31 27.681 & -02 05 19.66 & 5.78 & N
   & 11.49 & 9.91 & 8.46
   & - \\  

15 & 18 31 27.812 & -02 05 21.86 & 1.71 & N
   &  7.60  &  6.50  &  5.53 \tablenotemark{\dag}
   & IRS 1a \\  

16 & 18 31 28.016 & -02 05 17.90 & 0.94 & N
   & 10.81 & 9.44 & 8.56
   & - \\  

17 & 18 31 28.646 & -02 05 29.04 & 0.48 & N
   & - & - & - 
   & - \\  

18 & 18 31 28.654 & -02 05 29.74 & 11.1 & N
   & 11.23 & 8.93 & 7.27
   & IRS 1b \\  

19 & 18 31 28.683 & -02 05 22.25 & 0.20 & N
   & - & - & - 
   & - \\  

20 & 18 31 30.213 & -02 07 18.03 & 2.73 & Y($\sim$17)
   & - & - & - 
   & - \\  
\enddata
\tablenotetext{*}{From \citet{smi1985}.}
\tablenotetext{\dag}{Source was saturated in shortest exposure.  Magnitudes
quoted here are from 2MASS, translated to the MKO system.}

\end{deluxetable}
\end{small}

Of the 20 radio sources detected in our VLA data, 15 have counterparts
in our IR data with positions matched to within 1\arcsec.  In Table
\ref{tab:vlasources} we list the positions and flux densities of these
20 radio sources, averaged over our two observations.  We also note in
column 5 if they were found to be radio variable or not, by comparing
the observations of the two epochs and restricting the (u,v) coverage
of the images for comparison to the range of 16 to 300 k$\lambda$, a
coverage in common for the A and B configurations. The number given in
parentheses for the variable sources is the ratio between the largest
and smallest flux density observed. Columns 6,7, and 8 provide the
measured $J$, $H$, and $K$ magnitudes of the IR counterparts, if
any. As noted above, the four brightest IR counterparts (VLA-1,7,13,
and 15) were saturated in our data. For these sources we have
translated the published magnitudes in the 2MASS catalog to the MKO
system using the same transformation equations employed for our
calibration procedure. Typical internal errors on our measured IR
values are $<$0.1 magnitudes.

\section{Overall Characteristics of the Sample of Radio Sources}

\begin{sidewaysfigure*}[p]
  \centering
  \includegraphics[width=\textwidth,draft=false]{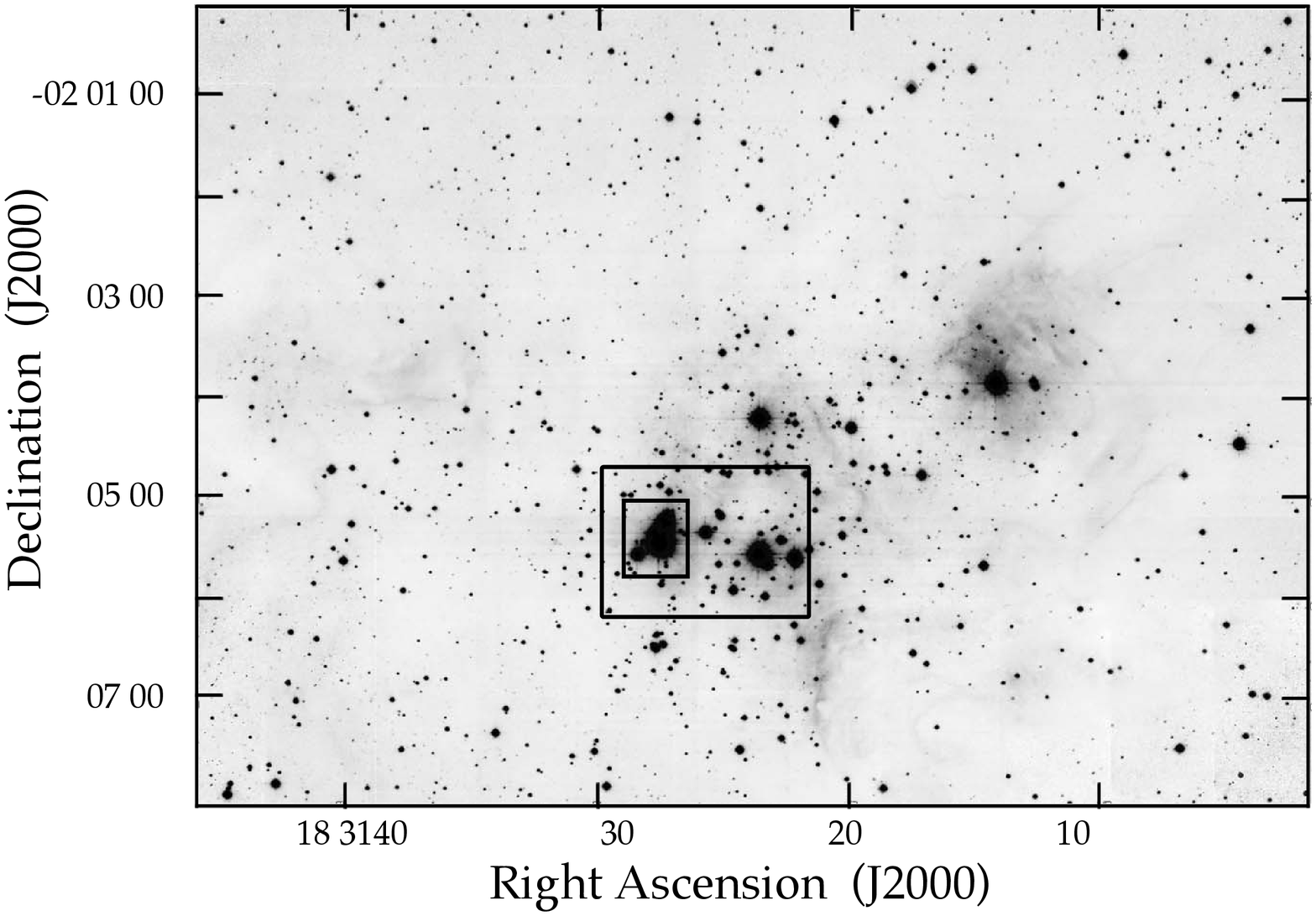}
  \caption{Composite $JHK^\prime$ image of the W40 region.  Two rectangles delineate the fields shown in Figures \ref{fig:wide_maps} and \ref{fig:zoom_maps}. }
  \label{fig:irmap_boxes}
\end{sidewaysfigure*}

\begin{figure*}[tbp]
  \centering
  \includegraphics[width=0.8\textwidth,draft=false]{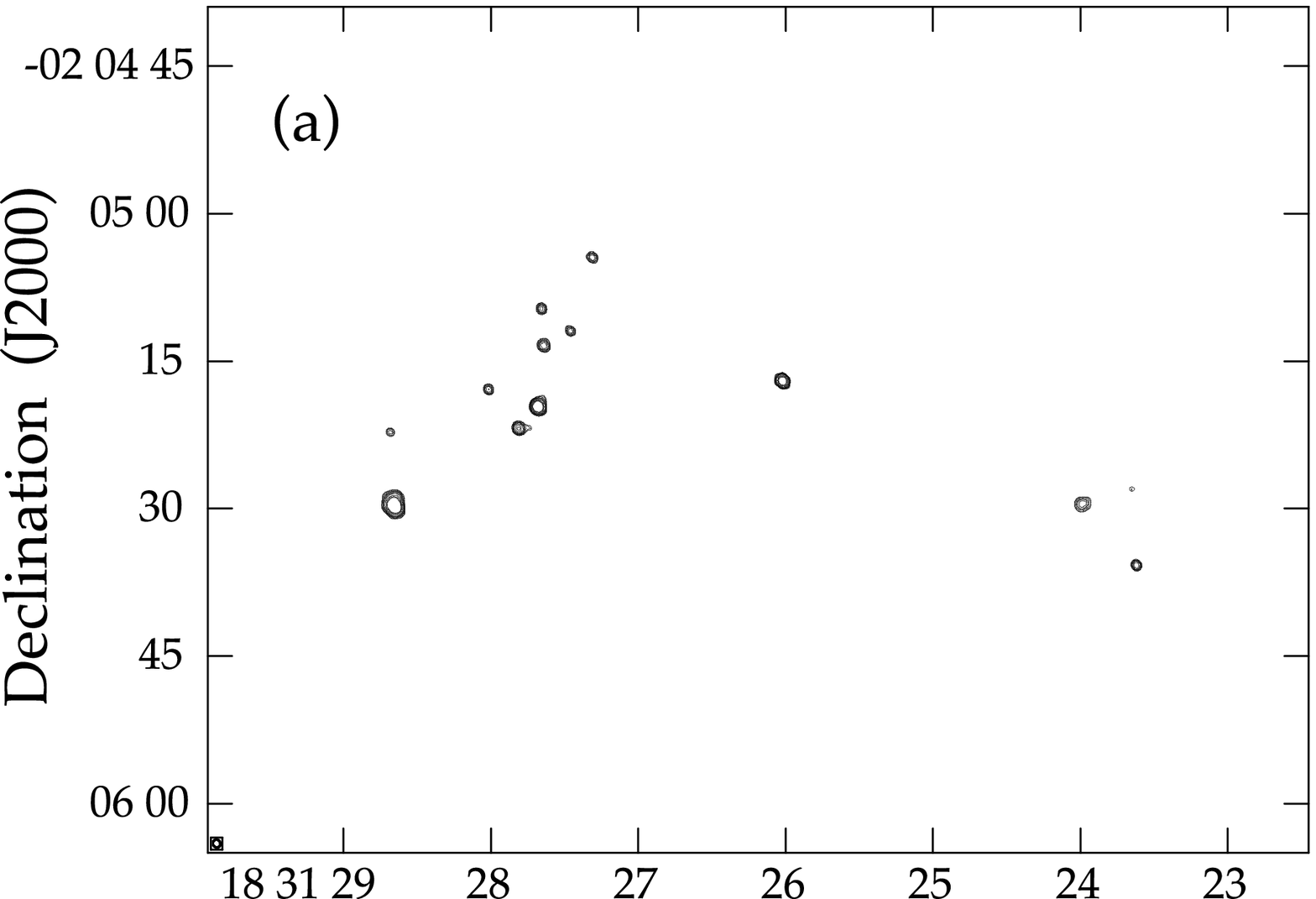}\vspace{3em}
  \includegraphics[width=0.8\textwidth,draft=false]{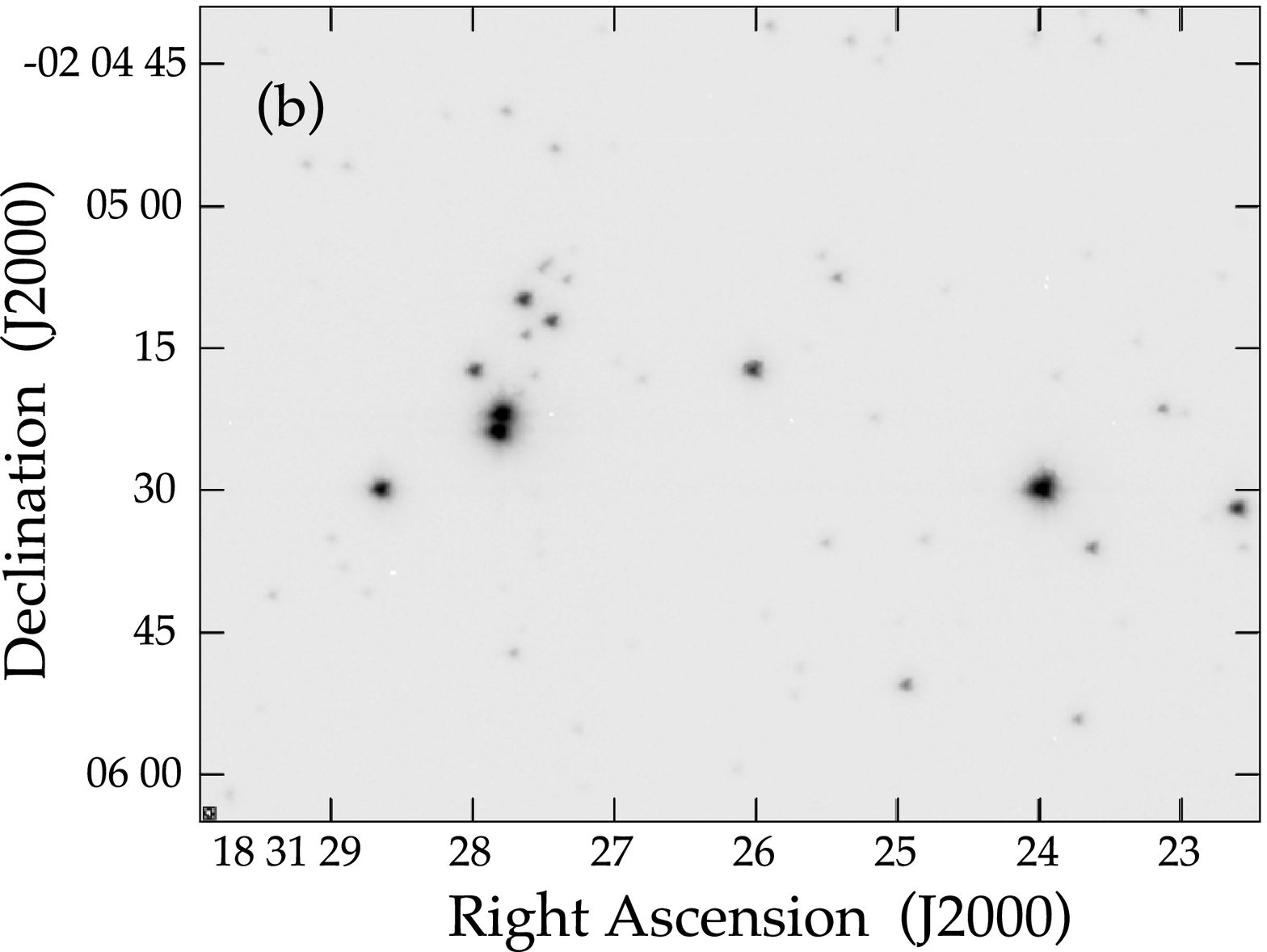}
  \caption{(a) VLA continuum image of the compact radio cluster in W40 at 3.6 cm. 
  The half power contour of the beam is shown in the bottom left corner. The contours 
  shown are 5, 6, 8, 10, 20, 40, 80, and 160 times 21   $\mu$Jy beam$^{-1}$, the rms 
  noise of the image.  (b) Composite IR map of the same region, showing that most of 
  the bright radio sources have IR counterparts.}
 \label{fig:wide_maps}
\end{figure*}

\begin{figure*}[tbp]
  \centering
  \includegraphics[width=0.6\textwidth,draft=false]{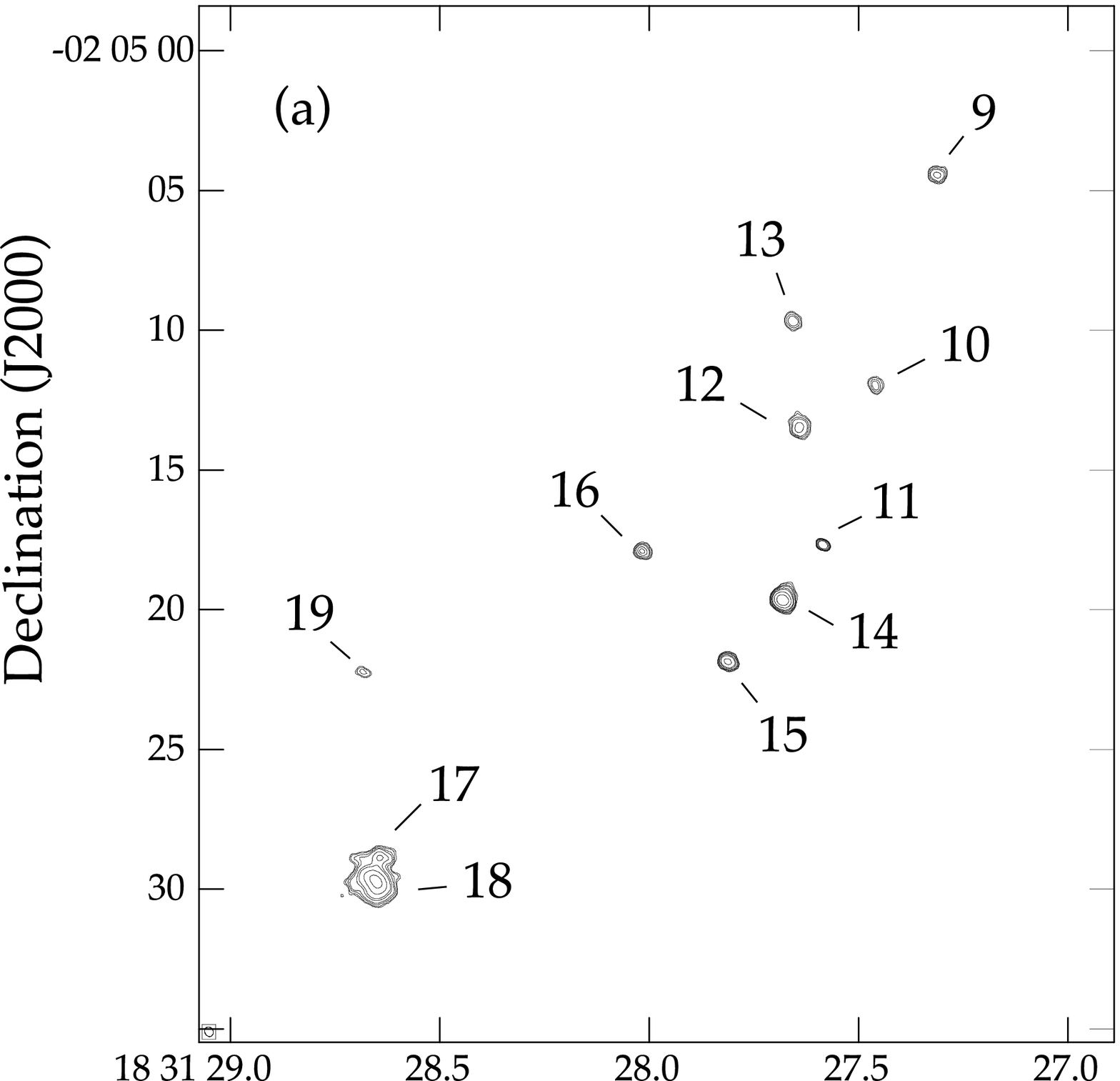}\vspace{3em}
  \includegraphics[width=0.6\textwidth,draft=false]{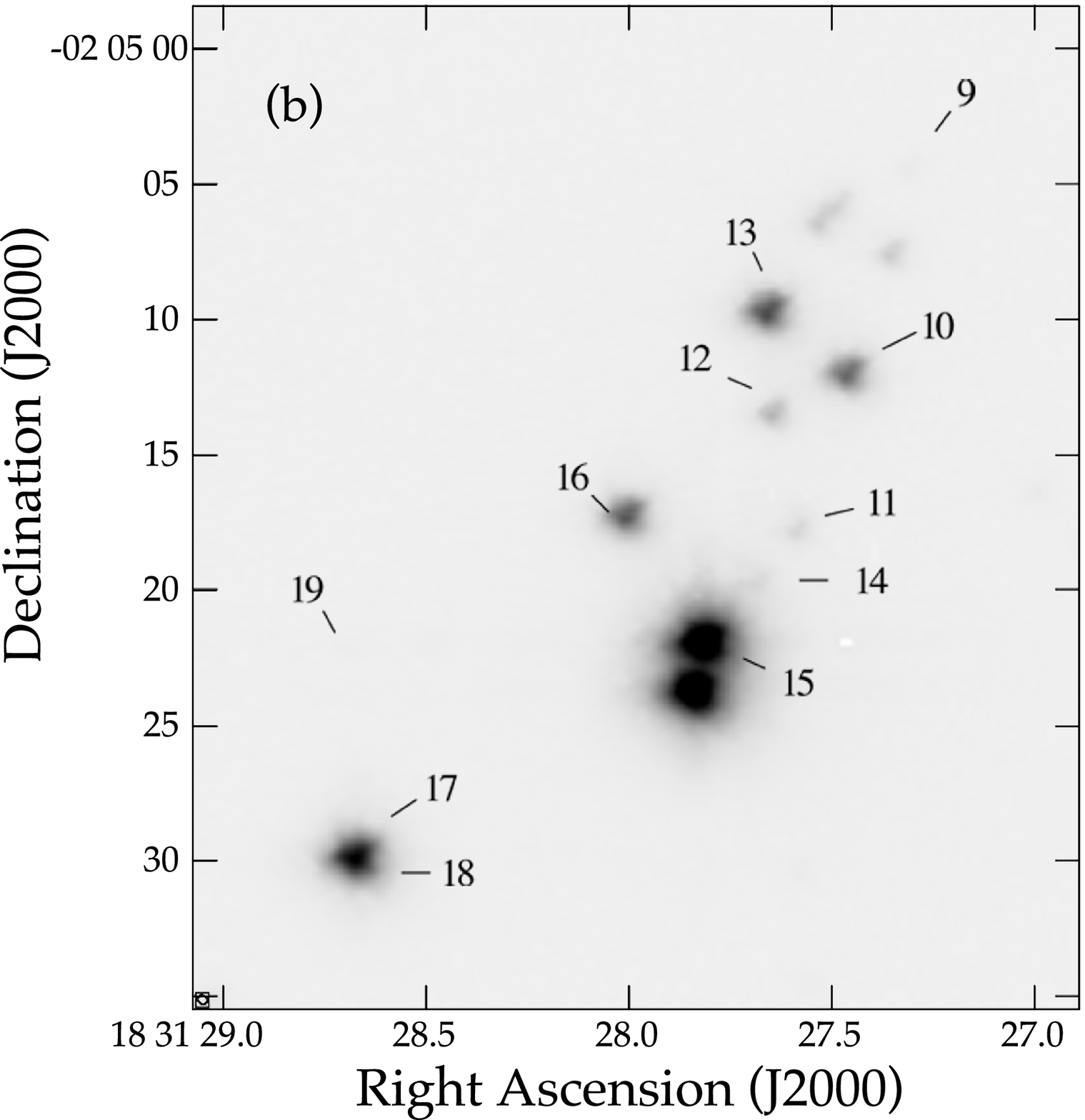}
  \caption{
  (a) Radio contour map as in Fig. \ref{fig:wide_maps}a.  This close-up view of the central 
  region shows 11 of the 20 radio sources.
  (b) Composite IR map covering the same region shown in (a). }
    \label{fig:zoom_maps}
\end{figure*}

A composite $JHK^\prime$ image of the entire W40 region is shown in Figure
\ref{fig:irmap_boxes}.  This field encompasses a region of $\sim 7\arcmin$
centered around the stellar cluster and the middle of the HII region. 
The two boxes in this figure indicate regions shown with greater detail in
Figures \ref{fig:wide_maps} and \ref{fig:zoom_maps}.  Figure \ref{fig:wide_maps}a
displays a radio contour map containing the entire radio cluster detected
at 3.6 cm. Out of the 20 detected radio sources, 11 of them $-$
including the 2 brightest $-$ are concentrated in a band with $30\arcsec$
of extent, as shown in Figure \ref{fig:zoom_maps}a.  
Figures \ref{fig:wide_maps}b and \ref{fig:zoom_maps}b show the same regions in the $K^\prime$ filter,
revealing that most of the radio sources can be associated with IR
counterparts.

Of the 20 sources detected, 10 were found to be time variable from one
observation to the other (see column 5 in Table 1).  This percentage (50\%)
of time variable sources is very similar to the value of 47\% found in Orion
by \citet{zap2004}. However, this percentage is larger than the value of 32\%
found in NGC 2024 by \citet{rod2003}. We searched unsuccessfully for circular
polarization in the sample. We measured the angular size of the sources using
the AIPS task IMFIT, with a correction for bandwidth smearing. With the
exception of sources VLA 8, VLA 13, and VLA 19, all sources were found to be
unresolved, $\theta_s \le 0$\farcs 2.

\section{Discussion}

There are several mechanisms that can produce compact centimeter radio
sources in regions of star formation.  In regions of low mass star formation,
thermal jets and gyrosynchrotron emitters can be present.  In regions of high
mass star formation we also have strong ionizing radiation available and, in
addition to the two mechanisms present in low mass star-forming regions, we
can have ionized stellar winds, ultracompact HII regions, and radio proplyds.
It is possible to distinguish between these various possibilities with high
angular resolution, multifrequency observations made with the required
sensitivity. 

Let us first consider the subset of 15 sources that have IR counterparts. It
is possible that these objects are ultracompact HII (UCHII) regions,
centered around young massive stars. The radio emission can then be explained
as optically thin free-free radiation, and the familiar photoionization
equilibrium equations \citep[e.g.][]{ost1974} can be used to determine physical
parameters of the system. However, UCHII regions are not expected
to exhibit short-period time variability, as is seen in 8 of the 15 radio/IR
sources. The remaining 7 sources can be considered as UCHII candidates.
Among this group are the two brightest radio sources, VLA-18 and VLA-14, as
well as the brightest IR source, VLA-15 = IRS1a.  With an assumed electron
temperature of $10^4$ K and a distance of 600 pc we can estimate that a B0
ZAMS star would be required to maintain the ionization in each of these UCHII
regions \citep{pan1973,tho1984}.

As for the sources that show time variability, these are more likely to
correspond to young stellar objects (YSOs) with active magnetospheres.  The
gyrosynchrotron radiation from such objects is expected to show substantial
variation over timescales of a few days.  One of the brightest radio/IR
sources that may be placed in this category is VLA-7 = IRS 2a. The SED of
this object has been examined from near-IR to millimeter wavelengths, and it
indicates that the star is surrounded by approximately 0.1 M$_{\sun}$ of 
circumstellar material \citep{val1994}.  This finding is consistent with our
tentative classification of VLA-7 as a YSO, since many stars remain shrouded
in their natal envelopes all the way to the main sequence. It is important to note
that this classification is at odds with the results of \citet{smi1985} and \citet{val1991},
who label this source as a deeply embedded OB star that is powering the ionization
of the W40 HII region.  Based on the data presented here, the radio sources VLA-14, 
15, and 18 may be equally valid candidates for supporting the HII region.

Of the five radio sources that do not have IR counterparts, VLA-17 and 19 are
of particular interest.  VLA-19 is offset by $7\farcs 5$ (0.02 pc) to the
north of VLA-18, and VLA-17 is only $0\farcs 7$ (0.002 pc) away to the
northwest.  These radio sources 
are both relatively weak, do not vary with time on short periods, and have
mildly non-circular radio contours. All of
this evidence suggests that these two radio sources may correspond to shock
fronts from a thermal jet interacting with the ambient interstellar medium. 
Based on its position and proximity, VLA-18 = IRS1b could well be the star
which powers this radio jet.

In order to unambiguously determine the nature of these compact radio
sources, further information is required.  A good indicator of UCHII
regions is the spectral index at radio frequencies.  Free-free
emission from an UCHII region should have a very flat radio spectrum,
so a spectral index more negative than -0.1 is a clear indication that
the object is not an UCHII region \citep{rod1993}.  This measurement
could be made using the EVLA to observe W40 with similar angular
resolution at 6 cm, for example.  Another important observation to
further our understanding of the W40 cluster will be to gather X-ray
data to determine the high-energy characteristics of the dominant
radio/IR sources.  Gyrosynchrotron emitters should produce detectable
X-ray emission, with measurable time-variability signatures.  A
combination of high angular resolution radio observations at longer
wavelengths plus X-ray measurements with XMM or Chandra would enable
us to definitively determine the astrophysical nature of these
sources.  A first Chandra study of W40 is currently in preparation
\citep{kuh2010}.

\section{Conclusions}

We have carried out the first detailed, high resolution study of the
center of the W40 region at 3.6 cm, supplemented by deep IR imaging of
the entire W40 complex.  Previous radio observations of W40 were not
able to distinguish individual sources, but our data have revealed a
cluster of some 15-20 radio sources.  Many of these sources are
clustered together in a narrow band at the center of the W40 cluster.
A substantial fraction of these radio objects are also present in our
$JHK^\prime$ images, and a number of them show significant variability
over timescales of a few days to weeks.  The combination of this high
resolution radio data and our IR maps allows us to speculate on the
nature of these sources, possibly distinguishing between YSOs, UCHII
regions, and shocked interstellar gas.

W40 is a rich, nearby star forming region that has not benefited from
the same intense scrutiny afforded to regions such as Orion,
Ophiuchus, and others.  This is largely due to the veil of gas and
dust which hides most of W40's compact objects from view.  Radio and
IR observations such as those described here are necessary to
penetrate the obscuring shell to investigate the interior.  X-ray
observations and further high resolution radio measurements at longer
wavelengths will be very beneficial in unmasking the W40 cluster.

\acknowledgements{ This material is based upon work supported by the
  National Aeronautics and Space Administration through the NASA
  Astrobiology Institute under Cooperative Agreement No. NNA08DA77A
  issued through the Office of Space Science.  This research has made
  use of the SIMBAD database, operated at CDS, Strasbourg, France, as
  well as NASA's Astrophysics Data System Bibliographic Services, and
  the NASA/ IPAC Infrared Science Archive, which is operated by the
  Jet Propulsion Laboratory, California Institute of Technology, under
  contract with the National Aeronautics and Space Administration.
  The infrared data presented here were obtained with the UH 2.2m
  telescope on Mauna Kea, operated by the University of Hawai'i. The
  authors wish to recognize and acknowledge the very significant
  cultural role and reverence that the summit of Mauna Kea has always
  had within the indigenous Hawaiian community.  We are most fortunate
  to have the opportunity to conduct observations from this mountain.
  Radio data presented here were obtained using the Very Large Array
  of the National Radio Observatory, a facility of the National
  Science Foundation operated under cooperative agreement by
  Associated Universities, Inc.}

\bibliographystyle{apj}
\bibliography{w40}

\begin{small}
\begin{deluxetable}{ccc r@{~$\pm$~}l r@{~$\pm$~}l r@{~$\pm$~}l} 
\label{tab:irsources}
\tablecolumns{9}
\tablecaption{Extended Photometry Table of W40 IR Sources (Electronic Version Only)}

\tablehead{ \colhead{Index}
	  & \colhead{$\alpha_{J2000}$}
	  & \colhead{$\delta_{J2000}$}
	  & \multicolumn{2}{c}{J}
	  & \multicolumn{2}{c}{H}
	  & \multicolumn{2}{c}{K}
	  }
\startdata

001 &  277.762394 &  -2.031111 &   16.608 &  0.145 &   14.838 &  0.203 &   13.809 &  0.272\\
002 &  277.764807 &  -2.005333 &   14.185 &  0.144 &   13.045 &  0.203 &   10.861 &  0.272\\
003 &  277.767017 &  -2.136254 &   15.328 &  0.144 &   12.772 &  0.203 &   11.734 &  0.272\\
004 &  277.767545 &  -2.181531 &   13.621 &  0.144 &   10.963 &  0.203 &    9.666 &  0.272\\
005 &  277.767801 &  -2.154991 &   15.387 &  0.144 &   13.509 &  0.204 &   12.394 &  0.272\\
006 &  277.767999 &  -2.010032 &   13.318 &  0.144 &   13.530 &  0.203 &   11.754 &  0.272\\
007 &  277.769254 &  -2.033301 &   16.745 &  0.146 &   14.706 &  0.203 &   13.349 &  0.272\\
008 &  277.770824 &  -2.054937 &   10.947 &  0.144 &   11.471 &  0.203 &   10.242 &  0.272\\
009 &  277.770948 &  -2.046558 &   13.686 &  0.144 &   12.726 &  0.203 &   11.683 &  0.272\\
010 &  277.771923 &  -2.109518 &   14.602 &  0.145 &   13.211 &  0.203 &   12.122 &  0.273\\
011 &  277.772396 &  -2.073557 &   10.221 &  0.144 &   10.612 &  0.203 &    8.327 &  0.272\\[1cm]
\multicolumn{9}{c}{ \parbox{0.8\tablewidth}{(complete table will be available in electronic version of AJ.  Contact authors directly for access before publication.)} }\\[1cm]

\enddata

\end{deluxetable}
\end{small}

\end{document}